 \documentclass[final,5p,times,twocolumn,authoryear]{elsarticle}

\usepackage{amssymb}
\usepackage{lipsum}
\usepackage{booktabs}

\journal{--}

\begin{document}

\begin{frontmatter}



\title{Ti substitution on Fe sites significantly changes the electronic properties of orthorhombic \(LaFeO_3\) perovskites — A first-principles study}  


\author[label1]{Jesaya Situmeang}
\author[label1]{Djoko Triyono}
\author[label1]{Muhammad Aziz Majidi}
\affiliation[label1]{organization={Department of Physics, University of Indonesia},
            city={Depok},
            country={Indonesia}}

\begin{abstract}
A large number of published experimental works suggest that when the Fe ions in orhorhombic \(LaFeO_3\) are substituted, band gap reduction is expected. However, recent experimental works observe band gap enhancement with increasing Ti ions replacing Fe ions. While satisfactory explanation on such observations seem absent, a first principles investigation may answer what should really happen.  We investigate from first-principles the influence of Ti-substitution on $LaFeO_3$ at Fe-site as a function of substitution concentration. Amongst the five investigated models, we found that as the Ti-substitution concentration increases, the electronic band gap at Fermi level decreases. However, in the model where two Ti ions replace Fe sites in an anti-symmetric arrangement, the Fermi level is crossed. We found that band gap reductions could be caused by the decreased in field splitting between the Fe 3d orbitals and charge competition between \(Fe-O\)  and \(Ti-O\) bonds as inferred from density of states analysis. While band gap reduction with increasing substitution implies better conductivity, cohesive energy becomes less negative although the perovskite distortion parameter does not differ significantly between each models.
\end{abstract}




\end{frontmatter}




\section{Introduction}
\label{introduction}
Orhthorhombic $LaFeO_3$ (LFO) perovskites posses advantageous properties for applications in humidity sensors, solar fuel cells, among others.
LFO properties can be further improved through various means, notably through substitutions at the cationic sites, either La or Fe.  Substitution at La sites have been thoroughly investigated, both experimentally and through first-principles.

Recently, Substitutions at Fe-sites have been one of the common ways to improve LFO properties [\cite{Triyono2020}, \cite{Zhou2021}, \cite{Mn-Triyono2019}, \cite{Hamada2011}]. In [\cite{Hamada2011}], through first-principles, Fe-sites of LFO surfaces are replaced by precious metals \((LaFe_{1-x}M_xO_3, M = Pd, Rh, Pt)\) to study enhancement in catalytic performance. Most recently, by replacing one Fe ion with Nobium (Nb), conduction is significantly improved because the Fermi level is crossed [\cite{Zhou2021}]. 

However, other ions can also replace Fe ions, as have been demonstrated by a number of experimental studies [\cite{Triyono2020}, \cite{TRIYONO2020102995}, \cite{Mn-Triyono2019}]. In most cases, band gap decreases with increasing substitution and electrical properties are improved. For example, in the case of Mg substitution [\cite{TRIYONO2020102995}], band gap decreases because the presence of Mg ions introduce more oxygen vacancies. In another case of Mn substitution [\cite{Mn-Triyono2019}], the decrease in band gap is attributed to the bond length. However, an experimental work on Ti substitution [\cite{Rini2020}] suggests an increase in band gap although satisfactory explanation has not been provided.

The difference in trend at the Ti substitution prompts for detailed investigation. One way to implement such is through first-principles calculations. In this study, we model Ti substitution on Fe sites for four concentrations \((0.25, 0.50, 0.75, 1.00)\). For each of these models, electronic band gaps are investigated, as well as several structural properties. 

\section{Computational Methods}
We employed Density Functional Theory (DFT) within the pseudopotential framework of Quantum-ESPRESSO [\cite{QEGiannozzi_2009}] to investigate the optimized structure and electronic properties of LFTO. Valence and core electrons interactions are described using ultrasoft pseudopotentials [\cite{USPP-GARRITY2014446}] where the generalized gradient approximation (GGA) is adopted to estimate the exchange-correlation potential in conjunction with the Perdew-Burke-Ernzherhof (PBE) exchange correlation functional [\cite{GGA-PBEPhysRevLett.77.3865}]. We set kinetic energy cutoff at 70 Ry with Monkhorst-Pack [\cite{moknhortspack-PhysRevB.13.5188}] k-points at \(6 \times 6 \times 4\). The optimised lattice parameters can be found in table \ref{tab:lattice}.

\subsection{Hubbard correction}
Since it is well known that GGA underestimates band gap of transitional metals, we incorporate appropriate Hubbard values for Fe and Ti 3d orbitals. The Hubbard U correction for Fe 3d orbital is determined at pristine LFO structure while the Ti 3d orbital at pristine \(LaTiO_3\) (LTO) structure until both band gap at Fermi level and lattice parameter are in good settlement with experimentally-proven values [\cite{PUSHPA2013184}]. We found 4.6eV as appropriate for Fe 3d orbital while 1.4eV for the Ti 3d orbital.

\subsection{Substitution modelling}
At room temperature, LFO (\(x=0.00\)) posses an Orthorhombic structure with G-AFM magnetic structure for the Fe atoms [\cite{PeterlinNeumaier1986}, \cite{Koehler1957}]. At the other end of the series, LTO (\(x=1.00\)) has the same crystallographic symmetry and magnetic structure [\cite{Furukawa1997}]. As for defect modelling, we replace one, two, three, and all four Fe ions with Ti to model \(x=0.25, 0.50, 0.75 \& 1.00\) respectively. For each model, structural relaxations are performed with force tolerance set at \(10^{-4}\) Ry/Bohr and electronic self-consistent calculations threshold at \(10^{-5}\) Ry. 

\subsection{Isolated atoms energy}
To compare cohesive energy amongst each model, we need to find out energy of isolated atoms — La, Fe, O, and Ti. We place each atom in a cubic space of 12$\mathring{A}$ with \(6\times6\times6\) k-points and energy cutoff set at 70 Ry. Electronic self-consistent calculations threshold is set at \(10^{-5}\) Ry. We use Martyna-Tuckerman correction on the total energy and self-consistent field potential because of the default periodic boundary condition assumption[\cite{Martyna1999}].





\section{Results and Discussion}

\begin{figure}
    \centering
    \includegraphics[width=0.75\linewidth]{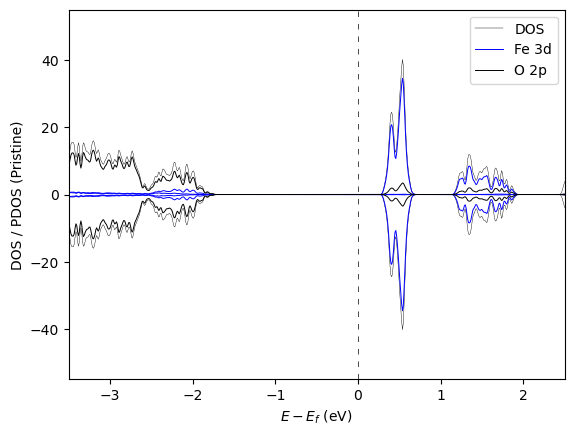}
    \caption{DOS/PDOS for Pristine \(LaFeO_3\). The energies are measured with respect to the Fermi level, indicated by the dashed vertical lines at \(E-E_f=0\). Band gap at Fermi level is calcuated as \(2.04eV\).}
    \label{fig:pristine-dos}
\end{figure}
\subsection{Electronic structure of pristine LFO}
Figure (\ref{fig:pristine-dos}) shows the density of states (DOS) of pristine LFO and projected density of states (PDOS) of orbitals Fe 3d, Ti 3d, and O 2p. We observe a bandgap of 2.04 eV which is still within the range of experimentally reported optical bandgap [\cite{TRIYONO2020102995}, \cite{Triyono2020}, \cite{TAGUCHI2005773}]. In addition, the plotted DOS is similar to a previously reported calculation [\cite{Scafetta_2014}] and valence orbitals near the Fermi level are consistent with reported photoelectron spectroscopy (PES) [\cite{Wadati2005}].

The shown occupied valence region lies between -3.5 eV to -1.77 eV. From -3.50 eV to -2.62 eV, O 2p orbitals overlaps with Fe 3d orbitals, indicative of strong hybridization. Within the -2.62 eV to -1.77 eV valence region, O 2p and Fe 3d orbitals have almost the same amount of peaks which implies significant \(Fe-O\) covalent bonding. The shown unoccupied states above the Fermi level can be divided into two regions — the first region from 0.23 eV to 0.69 eV; second region from 1.15 eV to 2 eV. Other than the band gap at Fermi level, we also observe a gap between the first and second region. While the band gap can be attributed to hybridization between the Fe 3d and O 2p orbitals [\cite{Rini2020}], the gap at conduction band is caused by \(e_g - t_{2g}\) crystal-field splitting [\cite{PUSHPA2013184}, \cite{Scafetta_2014}]. 

\subsection{Electronic structure after substitutions}

\begin{figure}
    \centering
    \includegraphics[width=0.75\linewidth]{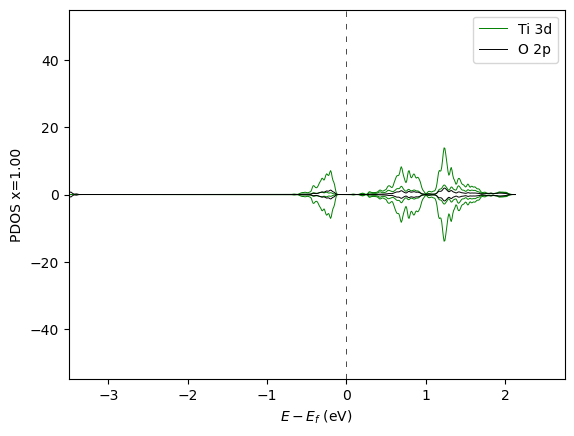}
    \caption{PDOS for \(x=1.00\) model or basically \(LaTiO_3\). Energies are measured with respect to the Fermi level, indicated by the dashed vertical line at \(E-E_f=0\). }
    \label{fig:lto}
\end{figure}

\begin{table}
    \centering
    \begin{tabular}{cc} \toprule
         \(x\)& \(E_{g}\) (eV)\\ \midrule
         0.00& 2.04\\ 
         0.25& 0.72\\ 
         0.50& —\\ 
         0.75& 0.25\\ 
         1.00&~0.2\\ \bottomrule
    \end{tabular}
    \caption{Summary of band gap values for investigated concentration. }
    \label{tab:bandgap}
\end{table}
\begin{figure}
    \centering
    \includegraphics[width=0.75\linewidth]{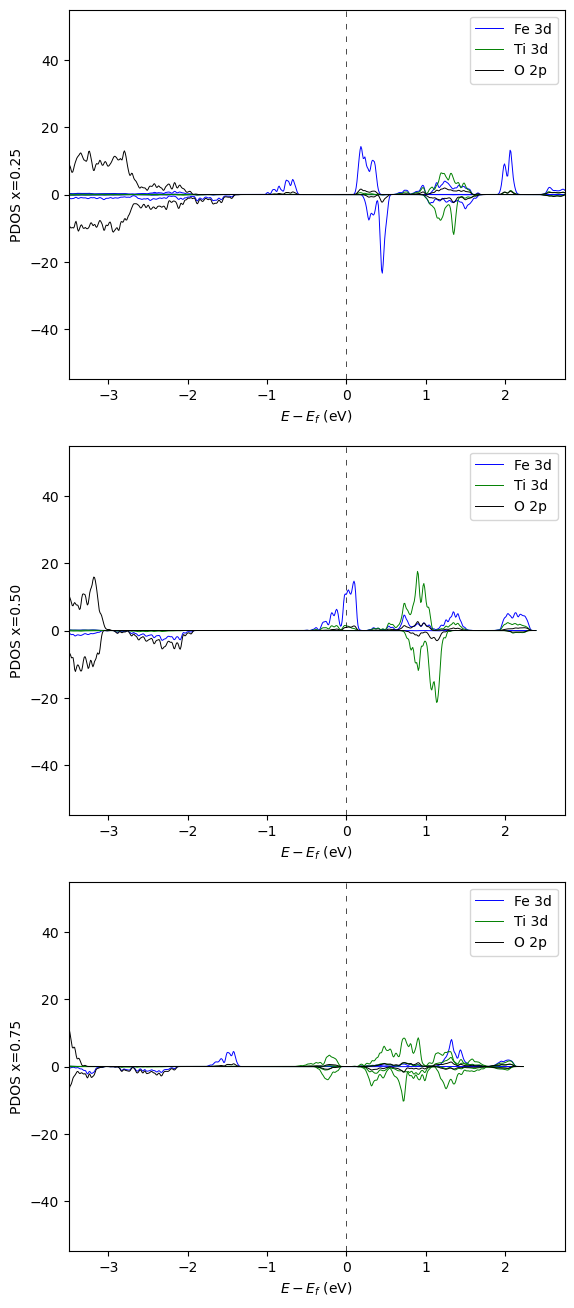}
    \caption{PDOS for \(x=0.25, 0.50, 0.75\) models. Energies are measured with respect to the Fermi level, indicated by the dashed vertical line at \(E-E_f=0\). } 
    \label{fig:0.25-0.75}
\end{figure}
Four Ti substitution on Fe site models are investigated. We found that for all models, band gap at Fermi level tend to decrease, except at \(x=0.50\), where the Fermi level is crossed. We will now discuss the PDOS curves illustrated in Figs. (\ref{fig:0.25-0.75}) and (\ref{fig:lto}) .

In the \(x=0.25\) model, Ti ion which substituted Fe ion has smaller charge and smaller significantly smaller magnetic moment.  As a result, crystal field splitting between Fe 3d \(e_g - t_{2g}\) orbitals near the Fermi level is reduced and hybridization between Fe 3d and O 2p orbitals are compromised, causing a decrease in band gap. Additionally, the introduction of Ti 3d orbitals formed impurity levels above the Fermi level, populating what used to be vacant in the pristine structure. A similar phenomena is observed within the \(x=0.75\) model. However, in the \(x=0.75\) model, the band gap at Fermi level is dominantly caused by field splitting between the Ti 3d \(e_g-t_{2g}\)orbitals, which is weaker than that of Fe, causing a much smaller band gap than the pristine model, and the band gap value at the \(x=1.00\) model confirm this dominance. Nevertheless, the presence of Fe orbitals above the Fermi level also populates what used to be vacant, improving overall conduction channels. At the \(x=1.00\) model (fig. ()), all of the Fe ions have been substituted by Ti ions, so that the produced DOS is similar to that of orthorhombic  \(LaTiO_3\) [\cite{Gu2016}].

Interestingly, within the \(x=0.50\) model, we found that there are no gaps at the Fermi level. To investigate further, we then plot charge density distributions for several models with isosurface level set at 0.07, as illustrated in figs. (). At models where gaps exist () charges still populate \(Fe-O\) bonds. However, with increasing Ti presence, charges from \(Fe-O\) bonds are seen to be pulled towards \(Ti-O\) bonds. This is consistent with the PDOS curves, where \(Fe-O\) hybridizations are reduced, and the bonding becomes less covalent. Meanwhile, at the \(x=0.50\) model, the presence of Ti pulls charges symmetrically so that there are no little-to-no charges populating the \(Fe-O\) bonds. 

While the charge distribution at all models agree on stronger covalent \(Ti-O\) bonds [\cite{Rini2020} ], there seems to be a correlation between the existence of band gaps with the degree of hybridization between \(Fe-O\). Within the \(x=0.50\) model, \(Ti-O\) hybridization wins over \(Fe-O\) due to \(Ti\) atomic positioning relative to \(Fe\). Since the hybridization between Fe 3d and O 2p orbitals plays a central role in forming the band gap of the pristine model, the diminutive presence of it would close the band gap.

\begin{figure}
    \centering
    \includegraphics[width=1\linewidth]{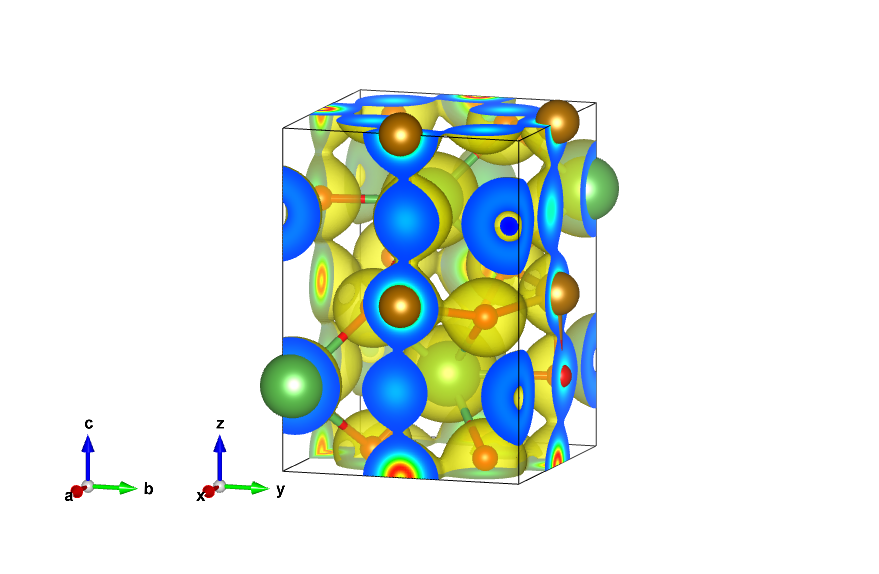}
    \caption{Charge density distribution of the pristine model. Brown-, blue-, red-, and green-colored balls refer to Fe, Ti, O, and La ions respectively.}
    \label{fig:pristine-charge}
\end{figure}
\begin{figure}
    \centering
    \includegraphics[width=1\linewidth]{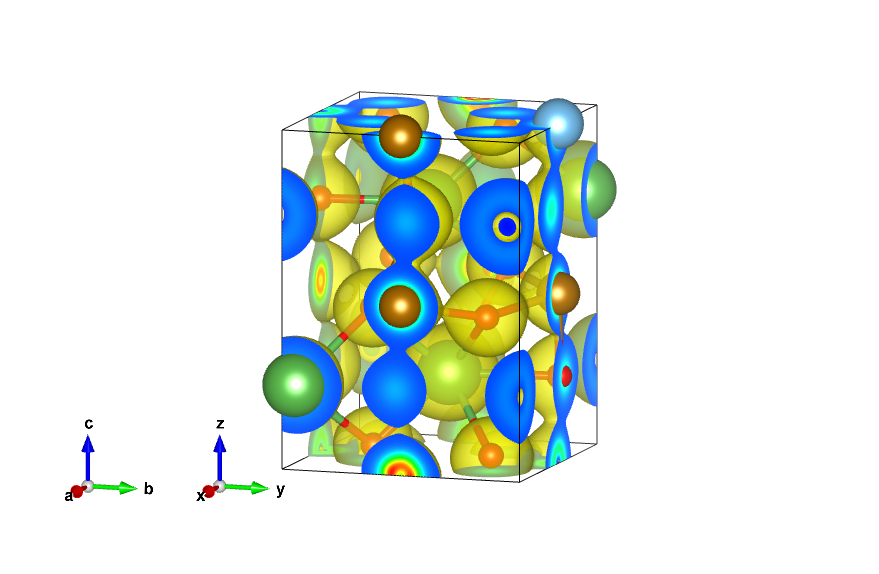}
    \caption{Charge density distribution of one Ti ion substitution model. The presence of Ti pulls charges from \(Fe-O\) bonds. Brown-, blue-, red-, and green-colored balls refer to Fe, Ti, O, and La ions respectively.}
    \label{fig:0.25 charge density}
\end{figure}
\begin{figure}
    \centering
    \includegraphics[width=1\linewidth]{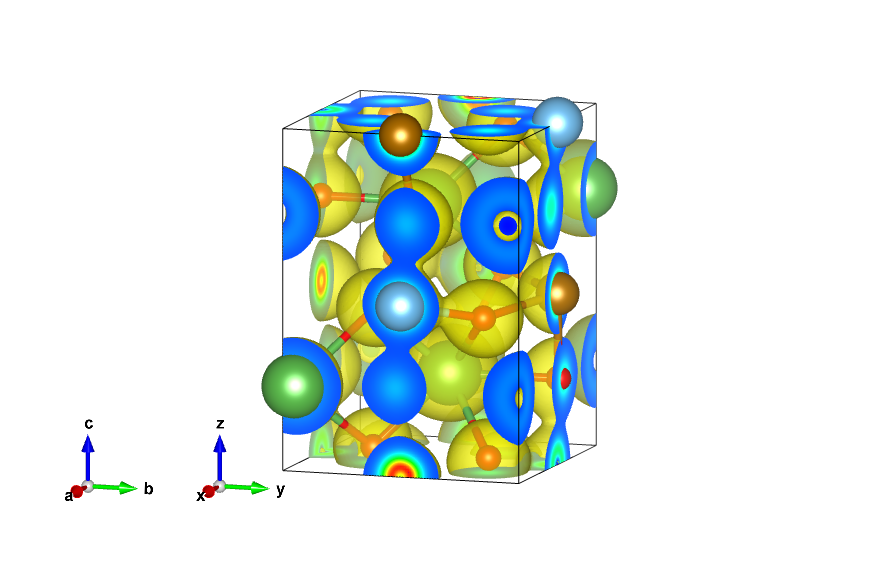}
    \caption{Charge density distribution of two Ti ions substitution model. In this model, the presence of Ti pulls charges from \(Fe-O\) bonds. Brown-, blue-, red-, and green-colored balls refer to Fe, Ti, O, and La ions respectively.}
    \label{fig:0.50 charge}
\end{figure}
\begin{figure}
    \centering
    \includegraphics[width=1.0\linewidth]{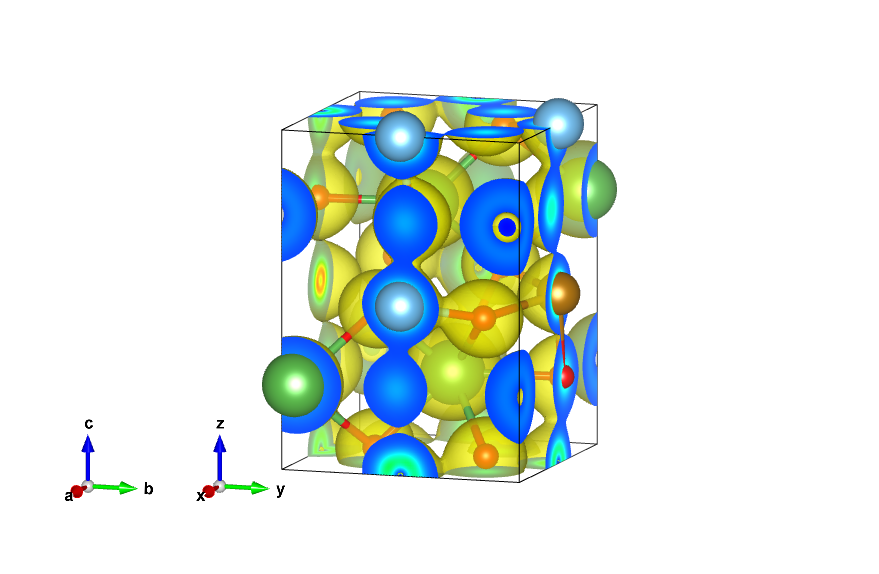}
    \caption{Charge density distribution of three Ti ions substitution model. In this model, the presence of Ti pulls charges from \(Fe-O\) bonds. Brown-, blue-, red-, and green-colored balls refer to Fe, Ti, O, and La ions respectively.}
    \label{fig:0.75 charge}
\end{figure}
\begin{figure}
    \centering
    \includegraphics[width=1\linewidth]{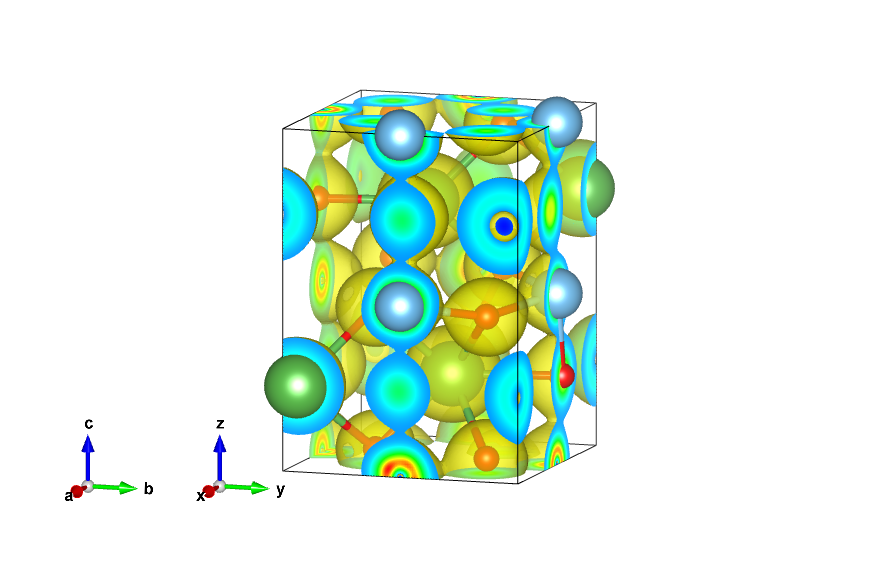}
    \caption{Charge density distribution of four Ti ions substitution model. In this model, the system no longer has Fe ions. Brown-, blue-, red-, and green-colored balls refer to Fe, Ti, O, and La ions respectively.}
    \label{fig:enter-label}
\end{figure}
\subsection{Structural stability}
To see how the substitutions affect structural stability, we compute cohesive energy for each model, using equation \ref{eq:cohesive}

\begin{equation} \label{eq:cohesive}
E_{chv} = E_{LaFe_{1-x}Ti_{x}O_3}-(E_{La}+(1-x)E_{Fe}+xE_{Ti}+3E_{O}).
\end{equation}

Values for the cohesive energies are tabulated in table \ref{tab:distortion} while optimised lattice parameters for each model can be seen in \ref{tab:lattice}. We have defined \(a_o=\sqrt[3]{\frac{V}{4}}\), \(V\) being the unit cell volume so that distortions are parametrized by the following equations, as proposed by ref. [\cite{Fang2006}]:

\[\alpha=1-\frac{a_o\sqrt{2}}{a},\]

\[\beta=1-\frac{a_o\sqrt{2}}{b},\]
and 
\[\gamma=1-\frac{2a_o}{c}.\]

\begin{table}
    \centering
    \begin{tabular}{cccc} \toprule
         \( x \)&  \( a \)(\AA) &  \( b \) (\AA) & \( c \) (\AA) \\ \midrule
         0&  5.5884&  5.6656& 7.9278
\\
         0.25&  5.6137&  5.7403& 7.8842
\\
         0.50&  5.6087&  5.7033& 7.9472
\\
         0.75&  5.6502&  5.7411& 7.8859
\\
         1.00&  5.6633&  5.6623& 7.9339
\\ \bottomrule
    \end{tabular}
    \caption{Calculated lattice parameters for each model.}
    \label{tab:lattice}
\end{table}

\begin{table}
    \centering
    \begin{tabular}{ccccc} \toprule
         \(x\)&  \( \alpha \)&  \( \beta \)&  \( \gamma \)& \( E_{chv} (Ry)\)\\ \midrule
         0&  0.9003&  0.8745&  0.8945
& -1352.8
\\ 
         0.25&  0.8994&  0.8576&  0.9126
& -1254.9
\\
         0.50&  0.9015&  0.8700&  0.8978
& -1156.9
\\
         0.75&  0.8915&  0.8615&  0.9166
& -1058.9
\\
         1.00&  0.8837&  0.8840&  0.9015
& -960.8
\\ \bottomrule
    \end{tabular}
    \caption{Calculated distortion and cohesive energy for each model.}
    \label{tab:distortion}
\end{table}

As the substitution concentration increases, cohesive energy becomes less negative while lattice parameter increases.  However, distortion does not differ significantly between models. This implies that with a lesser stability, the distortions within the perovskites are still maintained.

\section{Summary and conclusions}
Ti substitutions of Fe-sites within orthorhombic \(LaFeO_3\) have been investigated within the terms of electronic and structural properties. Four models were proposed, culminating in \(LaFe_{1-x}Ti_xO_3, x=0.25, 0.50, 0.75, 1.00\). As Ti ions replace Fe ions, covalent bonding between Fe and O atoms gets reduced, as well as the field splitting among Fe 3d orbitals. Since hybridization between the 3d orbitals of Fe ions and 2p orbitals of O ions plays an important role band gap formation, the presence of Ti generally reduces band gap, and depending on Ti positions relative to Fe, the system can not have any band gap at all. While Ti substition improves overall conduction due to gap reduction, distortion between the perovskites models do not differ significantly.

\section*{Acknowledgements}
Jesaya Situmeang would thank the Centre for Independent Learning at the University of Indonesia (CIL-UI) for grant no. PPK-26/ UN2.CIL/PDP.03.00/2023. 

\appendix

\bibliographystyle{elsarticle-harv} 
\bibliography{main}






\end{document}